\documentclass[aps,prd,showpacs,showkeys,floatfix,nofootinbib,twocolumn, 
notitlepage,superscriptaddress]{revtex4-1}
\usepackage{natbib}
\usepackage{color}
\usepackage{graphicx}
\usepackage{epsfig}
\usepackage{bm}
\usepackage{amsthm}
\usepackage{amsfonts}
\usepackage{float}
\usepackage{amsmath,amssymb}
\usepackage{graphicx}
\usepackage[caption = false]{subfig}
\usepackage{hyperref}
\usepackage{cleveref}

\newcommand{\sNN}{$\sqrt s_{ NN}$}
\newcommand{\COne}{$C_{1}$}

\newcommand{\CThree}{$C_{3}$}

\newcommand{\COneTwo}{$C_{1}/C_{2}$}
\newcommand{\CThreeTwo}{$C_{3}/C_{2}$}
\newcommand{\CFourTwo}{$C_{4}/C_{2}$}
\newcommand{\netKL}{net-($K$+$\Lambda$)}
\newcommand{\Ss}{$S\sigma$}

\newcommand{\gev}{GeV/$\it c$}
\begin{document}

\title{Cumulants of net-strangeness multiplicity distributions at energies
    available at the BNL Relativistic Heavy Ion Collider}

\date{ \today }

\author{Changfeng Li}
\affiliation{\it Institute of Frontier and Interdisciplinary Science, 
Shandong University, Qingdao, Shandong, 266237, China}
%\affiliation{
%Key Laboratory of Particle Physics and Particle Irradiation,
%Shandong University, Qingdao, Shandong, 266237, China\\}

\author{Deeptak Biswas}
\affiliation{\it The Institute of Mathematical Sciences, a CI of Homi 
Bhabha National Institute, Chennai, 600113, India}
 
\author{Nihar Ranjan Sahoo}

\affiliation {\it Institute of Frontier and Interdisciplinary Science, 
Shandong University, Qingdao, Shandong, 266237, China}
\affiliation{
Key Laboratory of Particle Physics and Particle Irradiation,
Shandong University, Qingdao, Shandong, 266237, China\\}
\affiliation{
National Institute of Science Education and Research, HBNI, Jatni 752050, India\\}
%%%%%%%%%%%%%%%%%%%%%%%%%%%%%%%%%%%%%%%%%%%%%%%%%%%%%%%%%%%%%%%%%%%%%%%%
\begin{abstract}  
The higher-order cumulants of net-proton number, net-charge, and 
net-strangeness multiplicity distributions are widely studied to search 
for the quantum-chromodynamics critical point and extract the 
chemical freeze-out parameters in heavy-ion collisions. In this 
context, the event-by-event fluctuations of the net-strangeness 
multiplicity distributions play important roles in extracting the 
chemical freeze-out parameter in the strangeness sector. Due to 
having difficulties in detecting all strange hadrons event by event, 
the kaon ($K$) and lambda ($\Lambda$) particles serve as a proxy 
for the strangeness-related observables in heavy-ion collisions. We 
have studied the net-$K$, net-$\Lambda$, and \netKL\ multiplicity 
distributions and calculated their different order of cumulants using 
the ultrarelativistic quantum molecular dynamics model and hadron 
resonance gas calculation. To adequately account for the net-strangeness 
cumulants, it has been found that the inclusion of resonance decay 
contributions in $K$ and $\Lambda$ is necessary.  
 \end{abstract}
%%%%%%%%%%%%%%%%%%%%%%%%%%%%%%%%%%%%%%%%%%%%%%%%%%%%%%%%%%%%%%%%%%%%%%%%
%For fun, see PACS list at  https://www.aip.org/publishing/pacs/pacs-2010-regular-edition
%\pacs{47.15.-x}

\maketitle 
%%%%%%%%%%%%%%%%%%%%%%%%%%%%%%%%%%%%%%%%%%%%%%%%%%%%%%%%%%%%%%%%%%%%%%%%
\section{Introduction}
%%%%%%%%%%%%%%%%%%%%%%%%%%%%%%%%%%%%%%%%%%%%%%%%%%%%%%%%%%%%%%%%%%%%%%%%
A deconfined phase of quantum chromodynamics (QCD) matter---known as the 
quark-gluon plasma---is created by colliding heavy ions at the BNL 
Relativistic Heavy-Ion Collider (RHIC) and the CERN Large Hadron 
Collider (LHC). The main goal of the RHIC 
Beam Energy Scan (BES) program is to search for the QCD critical point 
and gauge the QCD phase diagram in temperature ($T$) and baryon 
chemical potential ($\mu_{\rm B}$) plane. In BES phase-I, the STAR 
experiment reported several measurements on the higher-order cumulants 
of net-charge~\cite{STAR:2014egu}, net-proton number \cite{STAR:2013gus, 
STAR:2020tga,STAR:2021fge}, net-kaon \cite{STAR:2017tfy}, and  
net-$\Lambda$~\cite{STAR:2020ddh} 
multiplicity distributions as well. These cumulants are associated with 
their respective conserved charge susceptibilities, and hence these are 
related to thermodynamic quantities, like $T$ and $\mu_{\rm B}$, in 
heavy-ion collisions.

%%%%%%%%%%%%%%%%%%%%%%%%%%%%%%%%%%%%%%%%%%%%%%%%%%%%%%%%%%%%%%%%%%%%%%%%
The net-proton number cumulant measurements are proposed as the proxy 
of net-baryon number susceptibility; hence, it is used to search for the 
location of the QCD critical point~\cite{Stephanov:2004wx} in heavy-ion 
collision. References \cite{STAR:2013gus, STAR:2020tga, STAR:2021fge} have 
reported several net-proton cumulant measurements. In addition, one can 
extract the chemical freeze-out (CFO) temperature and $\mu_{\rm B}$ in 
the heavy-ion collisions with the help of these 
cumulants~\cite{Bazavov:2012vg, 
Gupta:2020pjd, Sharma:2021ncc}.

%%%%%%%%%%%%%%%%%%%%%%%%%%%%%%%%%%%%%%%%%%%%%%%%%%%%%%%%%%%%%%%%%%%%%%%%
On the other hand, the cumulants of net-kaon and net-$\Lambda$ 
multiplicity distribution act as a proxy for the strangeness and help 
to extract the CFO parameters from the strangeness sector, especially 
strangeness chemical potential ($\mu_{\rm S}$) and the respective 
temperature. The standard practice is to extract these parameters using 
the strange hadron yields ~\cite{STAR:2017sal, STAR:2019bjj} and also 
the higher order cumulants of net-strangeness multiplicity 
distributions. The inclusion of different resonances in the thermal 
model fit influences the CFO parameter~\cite{Alba:2020jir}. The freeze-
out temperature increases with the inclusion of heavier hadrons~\cite{ 
STAR:2017sal, Mitra:2020gxa}. Furthermore, the strange meson freezes 
out earlier than lighter hadrons at the highest RHIC energy, as studied 
in Refs.\cite{Bellwied:2018tkc, Chen:2020zuw}. Hence it is important to 
know the other strange baryons' freeze-out temperature and their 
chemical potential in heavy-ion collisions.

%%%%%%%%%%%%%%%%%%%%%%%%%%%%%%%%%%%%%%%%%%%%%%%%%%%%%%%%%%%%%%%%%%%%%%%%
In this paper, we study the cumulants of net-kaon, net-$\Lambda$, and 
net-(kaon+$\Lambda$) multiplicity distributions using the 
ultrarelativistic quantum molecular dynamics (UrQMD) model and 
the centrality variation for the STAR energies. To estimate the 
degree of thermalization, we have compared the most central results 
from UrQMD with the hadron resonance gas (HRG) calculation. These 
results would set a baseline for the ongoing measurements in the STAR 
experiment. The CFO parameters extracted from the thermal model fit of 
the particle yields are used in the HRG calculations to match the 
net-kaon and net-$\Lambda$ higher-order cumulant data from the STAR 
results~\cite{STAR:2019bjj}. A one-to-one comparison between UrQMD and 
HRG results for the  net-(kaon+$\Lambda$) elucidates the necessity to 
consider the contribution of the feed-down from the decay of higher 
mass resonance into the thermal model to explain the net-strangeness 
observable at RHIC energies.

%%%%%%%%%%%%%%%%%%%%%%%%%%%%%%%%%%%%%%%%%%%%%%%%%%%%%%%%%%%%%%%%%%%%%%%%
This paper is organized as follows. In 
Sec.~\ref{Sect:cumulantIntro}, 
the definition of net-(kaon+$\Lambda$) and their cumulants are introduced. 
A brief introduction of the UrQMD model and HRG calculation are 
mentioned in Secs.~\ref{Sect:UrQMDIntro} and ~\ref{Sect:hrgintro}, 
respectively. The net-kaon, net-$\Lambda$, and net-(kaon+$\Lambda$) cumulants 
and their ratios are discussed in Sec.~\ref{Sect:urqmdDis}. The 
comparison between the UrQMD and HRG calculations for the net-$K$ and 
net-$\Lambda$ multiplicity distributions is discussed in 
Sec.~\ref{Sect:urqmdHrg}. The strangeness production and resonance 
decay effects are discussed in Sec.~\ref{Sect:hrgstrngeproduction}. 
Finally, we summarize our studies and outline the STAR measurements as 
an outlook in Sec.~\ref{Sect:summaryoutlook}.
%%%%%%%%%%%%%%%%%%%%%%%%%%%%%%%%%%%%%%%%%%%%%%%%%%%%%%%%%%%%%%%%%%%%%%%% 
\begin{figure*}[htb!]
    \centering
    \includegraphics[width=1.0\textwidth]{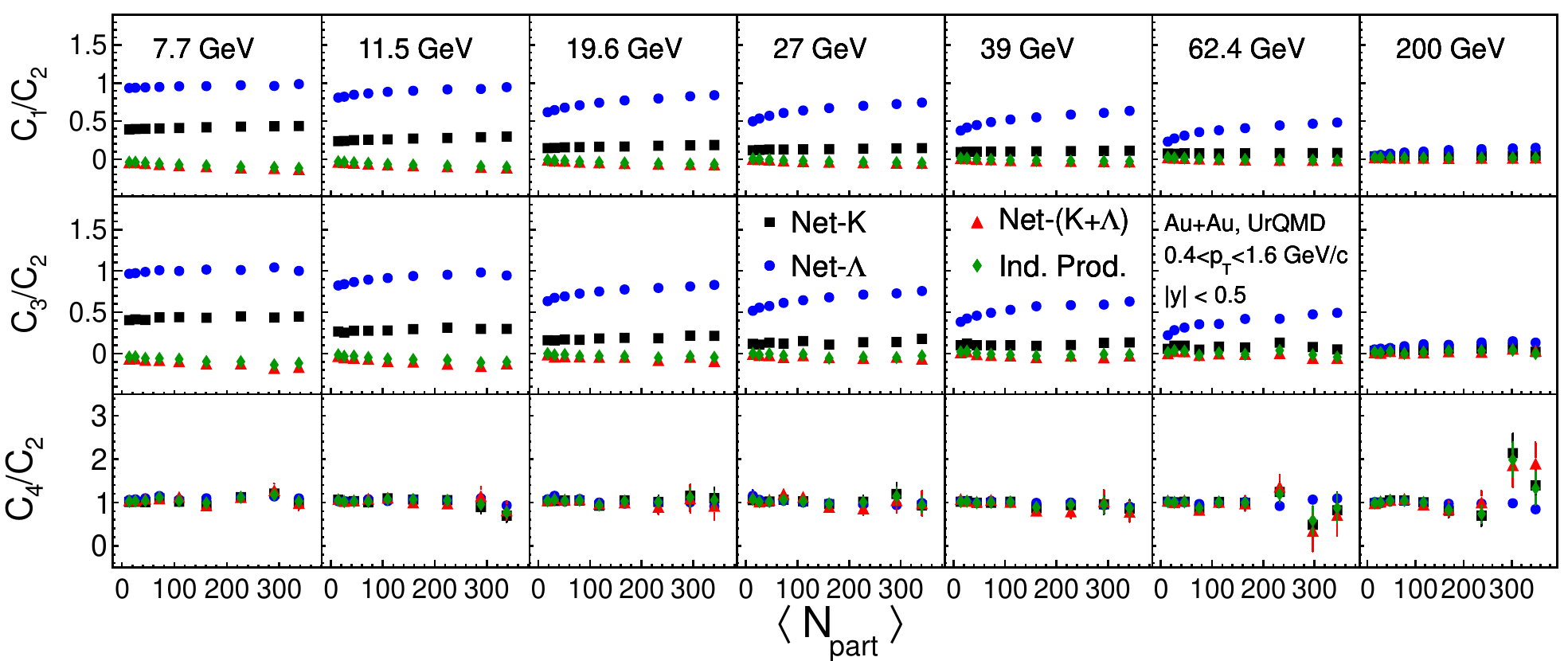}
    \caption{The cumulants ratios of net-$K$ (box), 
    net-$\Lambda$(circle), net-($K+\Lambda$) (triangle), and 
    independent particle production for net-($K+\Lambda$) (diamond)  are shown at \sNN=7.7--200 
    GeV calculated 
within the same acceptance. The vertical bars are statistical errors.}
    \label{fig:CumulnatsRatio}
\end{figure*}
%%%%%%%%%%%%%%%%%%%%%%%%%%%%%%%%%%%%%%%%%%%%%%%%%%%%%%%%%%%%%%%%%%%%%%%%

%%%%%%%%%%%%%%%%%%%%%%%%%%%%%%%%%%%%%%%%%%%%%%%%%%%%%%%%%%%%%%%%%%%%%%%%
\section{Observables and models}
%%%%%%%%%%%%%%%%%%%%%%%%%%%%%%%%%%%%%%%%%%%%%%%%%%%%%%%%%%%%%%%%%%%%%%%%
\subsection{Cumulants of Net-(kaon+$\Lambda$) multiplicity distributions}
\label{Sect:cumulantIntro}
%%%%%%%%%%%%%%%%%%%%%%%%%%%%%%%%%%%%%%%%%%%%%%%%%%%%%%%%%%%%%%%%%%%%%%%%
We define the number of net-(kaon+$\Lambda$) [in short \netKL] as
%%%%%%%%%%%%%%%%%%%%%%%%%%%%%%%%%%%%%%%%%%%%%%%%%%%%%%%%%%%%%%%%%%%%%%%%
\begin{equation}
N_{K+\Lambda} = (N_{K^{+}} + N_{\bar{\Lambda}}) - (N_{K^{-}} + 
N_{\Lambda}). 
\label{Eq:NetKL}
\end{equation}
%%%%%%%%%%%%%%%%%%%%%%%%%%%%%%%%%%%%%%%%%%%%%%%%%%%%%%%%%%%%%%%%%%%%%%%%
Here, $N_x$ is the number of $x$ particle in an event within a 
given phase space window, where $x$ is $K^{+} (u\bar{s}), K^{-} 
(\bar{u}s), \Lambda (uds)$, or $\bar{\Lambda} (\bar{u}\bar{d}\bar{s})$. 
Here, the $(N_{K^{+}} + N_{\bar{\Lambda}})$ and $(N_{K^{-}} + 
N_{\Lambda})$ represents the total positive and negative strangeness 
quantum numbers in an event, respectively. In heavy-ion collisions, the 
initial strangeness is zero. Note here that the STAR net-kaon and 
net-$\Lambda$ publications are reported with ($N_{K^{+}} - N_{K^{-}}$) in 
net-kaon ~\cite{STAR:2017tfy} and ($N_{\Lambda} - N_{\bar{\Lambda}}$) 
in net-$\Lambda$ measurements~\cite{STAR:2020ddh}, respectively. 

%%%%%%%%%%%%%%%%%%%%%%%%%%%%%%%%%%%%%%%%%%%%%%%%%%%%%%%%%%%%%%%%%%%%%%%%
The $\langle N \rangle$ is the ensemble average of multiplicity in a 
given centrality class. The deviation of $N$ from its mean 
$\langle N \rangle$ is defined as $\delta N = N -  \langle N  \rangle 
$. The $n$th order of diagonal cumulants ($C_{n}$) can be defined as
%%%%%%%%%%%%%%%%%%%%%%%%%%%%%%%%%%%%%%%%%%%%%%%%%%%%%%%%%%%%%%%%%%%%%%%%
\begin{align}
C_{1} &= \langle  N \rangle, \\
C_{2} &= \langle  (\delta N)^{2} \rangle\\
C_{3} &= \langle  (\delta N)^{3} \rangle\\
C_{4} &= \langle  (\delta N)^{4} \rangle - 3  \langle  (\delta N)^{2} \rangle^{2}
\end{align}
%%%%%%%%%%%%%%%%%%%%%%%%%%%%%%%%%%%%%%%%%%%%%%%%%%%%%%%%%%%%%%%%%%%%%%%%
A detailed discussion regarding the cumulant-generating function and 
relationships between moments, central moments, and cumulants can be 
found in Ref.~\cite{Kitazawa:2017ljq}. In this paper, we discuss 
different orders of cumulants and their ratios for the 
net-($K$+$\Lambda$), net-$K$, and net-($\Lambda$) multiplicity 
distributions. The connection between these cumulants and the 
thermodynamics susceptibilities is discussed in Sec.~\ref{Sect:hrgintro}. 

%%%%%%%%%%%%%%%%%%%%%%%%%%%%%%%%%%%%%%%%%%%%%%%%%%%%%%%%%%%%%%%%%%%%%%%%
\subsection{UrQMD model}
\label{Sect:UrQMDIntro}
%%%%%%%%%%%%%%%%%%%%%%%%%%%%%%%%%%%%%%%%%%%%%%%%%%%%%%%%%%%%%%%%%%%%%%%%
The UrQMD~\cite{Bleicher:1999xi, Bass:1998ca} model is a microscopic 
transport model. In this space-time evolution model, the propagation, 
rescattering among hadrons, and string excitation are included, 
whereas no in-medium modification effects are implemented. Hence, this 
model is used to study the baseline measurement in heavy-ion collisions 
for various observables~\cite{Luo:2014tga, Chatterjee:2016mve}.

%%%%%%%%%%%%%%%%%%%%%%%%%%%%%%%%%%%%%%%%%%%%%%%%%%%%%%%%%%%%%%%%%%%%%%%%
In the present work, the Au+Au collision events are simulated at 
\sNN= 7.7, 11.5, 19.6, 27, 39, 62.4, and 200 GeV with minimum-bias 
configuration (impact parameter, $b = 0$ to 14 fm). The different 
particles with their particle data group particle identification 
in an event are analyzed using the same phase space window similar to 
the STAR experiment. The centrality selections are performed based on 
the charged particle multiplicity distribution at the midrapidity, 
following the STAR experiment specifications. The net-$K$, net-$\Lambda$, 
and \netKL\ multiplicity distributions are calculated within $0.4 < 
p_{T}< 1.6$ \gev\ and rapidity window $|y|<$0.5. The cumulants and 
their ratios are calculated as mentioned in 
Sec.~\ref{Sect:cumulantIntro}. 
%%%%%%%%%%%%%%%%%%%%%%%%%%%%%%%%%%%%%%%%%%%%%%%%%%%%%%%%%%%%%%%%%%%%%%%%
\subsection{HRG calculation}
\label{Sect:hrgintro}
%%%%%%%%%%%%%%%%%%%%%%%%%%%%%%%%%%%%%%%%%%%%%%%%%%%%%%%%%%%%%%%%%%%%%%%%
The ideal HRG model considers an ensemble of noninteracting hadrons and 
their resonances. The logarithm of the partition function can be written 
as
%%%%%%%%%%%%%%%%%%%%%%%%%%%%%%%%%%%%%%%%%%%%%%%%%%%%%%%%%%%%%%%%%%%%%%%%
\begin{equation}
\ln Z^{\rm id}=\sum_{i} \ln Z_{i}^{\rm id},
\end{equation}
%%%%%%%%%%%%%%%%%%%%%%%%%%%%%%%%%%%%%%%%%%%%%%%%%%%%%%%%%%%%%%%%%%%%%%%%
where the sum runs over all the hadrons and resonances; the `$\rm id$' 
stands for the ideal gas consideration. The Boltzmann approximation 
provides a reasonable baseline for the massive hadrons and resonances 
(except $\pi$) along the chemical freeze-out boundary 
\cite{Wheaton:2004qb}. In this work, the particle species under 
consideration ($K, \Lambda$) have much higher masses than the freeze-out 
temperature and $m_i-\mu_i >> T$  for the respective freeze-out 
parametrization. Within this consideration, we can approximate the 
partition functions in the Boltzmann limit \cite{Kitazawa:2012at}. With 
such assumption \cite{Wheaton:2004qb},
%%%%%%%%%%%%%%%%%%%%%%%%%%%%%%%%%%%%%%%%%%%%%%%%%%%%%%%%%%%%%%%%%%%%%%%%
\begin{equation}\label{ZHRG}
\ln Z_{i}^{\rm id}=\pm \frac{V g_{i}}{2 \pi^{2}} \int_{0}^{\infty} p^{2} dp ~e^{ -\left(E_{i}-\mu_{i}\right) / T}
\end{equation}
%%%%%%%%%%%%%%%%%%%%%%%%%%%%%%%%%%%%%%%%%%%%%%%%%%%%%%%%%%%%%%%%%%%%%%%%
where $V$, $T$, and $g_{i}$  are the system volume, temperature, and 
degeneracy factor of the $i$th hadron. $E_{i}=$ 
$\sqrt{p^{2}+m_{i}^{2}}$ is the single particle energy. $\mu_{i}=B_{i} 
\mu_{B}+S_{i} \mu_{S}+Q_{i} \mu_{Q}$ is the chemical potential, where 
$B_{i}, S_{i},$ and $Q_{i}$ are, respectively, the baryon number, 
strangeness, and charge of the particle. The $\mu_{B}, \mu_{S}$, and 
$\mu_{Q}$ are the baryon, strangeness, and charge chemical potentials, 
respectively.  

%%%%%%%%%%%%%%%%%%%%%%%%%%%%%%%%%%%%%%%%%%%%%%%%%%%%%%%%%%%%%%%%%%%%%%%%
With the Boltzmann approximation, the pressure of a single hadron species 
$i$ is defined as \cite{Wheaton:2004qb, Huovinen:2017ogf} 
%%%%%%%%%%%%%%%%%%%%%%%%%%%%%%%%%%%%%%%%%%%%%%%%%%%%%%%%%%%%%%%%%%%%%%%%
\begin{eqnarray}
P_{i}^{\rm id}&=&\frac{T}{V} \ln Z_{i}^{\rm id} \nonumber  
     =\pm \frac{g_{i} T}{2 \pi^{2}} \int_{0}^{\infty} p^{2} 
     dp  ~e^{ -\left(E_{i}-\mu_{i}\right) / T} 
\end{eqnarray}
%%%%%%%%%%%%%%%%%%%%%%%%%%%%%%%%%%%%%%%%%%%%%%%%%%%%%%%%%%%%%%%%%%%%%%%%
The $n$th order susceptibility is defined as
%%%%%%%%%%%%%%%%%%%%%%%%%%%%%%%%%%%%%%%%%%%%%%%%%%%%%%%%%%%%%%%%%%%%%%%%
\begin{equation}
\chi_{x}^{n}=\frac{1}{V T^{3}} \frac{\partial^{n}(\ln Z)}{\partial\left(\frac{\mu_{x}}{T}\right)^{n}}
\end{equation}
%%%%%%%%%%%%%%%%%%%%%%%%%%%%%%%%%%%%%%%%%%%%%%%%%%%%%%%%%%%%%%%%%%%%%%%%
where $\mu_{x}$ is the chemical potential for conserved charge $x$. 
The derivatives of the logarithm of the grand canonical partition function 
$(Z)$ with respect to the chemical potential define susceptibilities.
These susceptibilities are related to the event-by-event measured 
cumulants of net-charge, net-baryon, and net-strangeness multiplicity 
distributions. For our present purpose $x=S$ 
(strangeness).

%%%%%%%%%%%%%%%%%%%%%%%%%%%%%%%%%%%%%%%%%%%%%%%%%%%%%%%%%%%%%%%%%%%%%%%%
With the Boltzmann limit in Eq.(\ref{ZHRG}), the $n$th order 
susceptibility can be expressed as, 
%%%%%%%%%%%%%%%%%%%%%%%%%%%%%%%%%%%%%%%%%%%%%%%%%%%%%%%%%%%%%%%%%%%%%%%%
\begin{eqnarray}\label{Eq.cumulant}
\chi_{x}^{n}&=&\sum_{i} \frac{g_{i} x^{n}_{i}}{2 \pi^{2} T^{3}} \int_{0}^{\infty} p^{2} d p ~e^{ -\left(E_{i}-\mu_{i}\right) / T}. \label{mom1}
\end{eqnarray}

%%%%%%%%%%%%%%%%%%%%%%%%%%%%%%%%%%%%%%%%%%%%%%%%%%%%%%%%%%%%%%%%%%%%%%%%
This form simplifies the representation of cumulants, making 
the multiplicity distribution similar to a Poisson distribution 
\cite{Kitazawa:2012at}. This simplification is due to the fact that the 
derivative of the exponential function is also exponential and the 
analytical form is independent of the order of derivative $n$. 
%For particles under consideration ($K$ and $\Lambda$), 
In Boltzmann limit, the relationship 
between different cumulants becomes straightforward with $C_1=C_3$ and 
$C_2=C_4$. Experimental observations at RHIC energies, as reported by the STAR experiment, confirm this Boltzmann approximation  by considering the Poisson distributions in the net-$K$ and net-$\Lambda$ cumulants measurements
\cite{STAR:2017tfy, STAR:2020ddh}.

%%%%%%%%%%%%%%%%%%%%%%%%%%%%%%%%%%%%%%%%%%%%%%%%%%%%%%%%%%%%%%%%%%%%%%%%
\subsection{Connection with experimental observable}
%%%%%%%%%%%%%%%%%%%%%%%%%%%%%%%%%%%%%%%%%%%%%%%%%%%%%%%%%%%%%%%%%%%%%%%%
The event-by-event net-charge multiplicity distributions are measured 
in heavy-ion experiments within a finite acceptance. The cumulants 
($C_{n}$) discussed in Sec.~\ref{Sect:cumulantIntro} are related to 
the different order of susceptibilities by the following relation:
%%%%%%%%%%%%%%%%%%%%%%%%%%%%%%%%%%%%%%%%%%%%%%%%%%%%%%%%%%%%%%%%%%%%%%%%
\begin{equation}
V T^{3}\chi_{x}^{n}= C_n .
\end{equation}
%%%%%%%%%%%%%%%%%%%%%%%%%%%%%%%%%%%%%%%%%%%%%%%%%%%%%%%%%%%%%%%%%%%%%%%%
The ratios of these cumulants are taken to cancel the volume term in 
the above expression. The mean $\left(M_{q}\right),$ variance 
$\left(\sigma_{q}^{2}\right),$ skewness $\left(S_{q}\right)$, and 
kurtosis $\left(\kappa_{q}\right)$ are also related with different 
cumulants as follows: 
%%%%%%%%%%%%%%%%%%%%%%%%%%%%%%%%%%%%%%%%%%%%%%%%%%%%%%%%%%%%%%%%%%%%%%%%
\begin{eqnarray}
\sigma_{q}^{2} / M_{q} &=& C_{2} / C_{1}=\chi_{q}^{2} / \chi_{q}^{1} \\
S_{q} \sigma_{q} &=& C_{3} / C_{2}=\chi_{q}^{3} / \chi_{q}^{2} \\
\kappa_{q} \sigma_{q}^{2} &=& C_{4} / C_{2}=\chi_{q}^{4} / \chi_{q}^{2}
\end{eqnarray}

%%%%%%%%%%%%%%%%%%%%%%%%%%%%%%%%%%%%%%%%%%%%%%%%%%%%%%%%%%%%%%%%%%%%%%%%
\section{Results and discussion}
%%%%%%%%%%%%%%%%%%%%%%%%%%%%%%%%%%%%%%%%%%%%%%%%%%%%%%%%%%%%%%%%%%%%%%%%
\subsection{Cumulants of Net-$K$, net-$\Lambda$, and \netKL\ in UrQMD}
\label{Sect:urqmdDis}
%%%%%%%%%%%%%%%%%%%%%%%%%%%%%%%%%%%%%%%%%%%%%%%%%%%%%%%%%%%%%%%%%%%%%%%%
The $K^{-}/K^{+}$ yield ratio increases with collision energy and 
approaches to unity at higher collision energy~\cite{STAR:2017sal} due 
to the interplay between associated production ($NN \rightarrow KYN$, 
$\pi N \rightarrow KY$) and pair production ($NN \rightarrow 
NN K^{+}K^{-}$). The $\bar{\Lambda}/\Lambda$ yield ratio also shows the 
same trend as a function of collision energy~\cite{Andronic:2005yp}. 
Recent net-$K$~\cite{STAR:2017tfy} and net-$\Lambda$~\cite{STAR:2020ddh} 
measurements show that the \COne\ of these multiplicity distributions 
are positive and increase with centrality and also collision energy. In 
this paper, we calculate the cumulant ratios of net-$K$, net-$\Lambda$, 
and \netKL\ multiplicity distributions for seven collision energies 
using the UrQMD model.

%%%%%%%%%%%%%%%%%%%%%%%%%%%%%%%%%%%%%%%%%%%%%%%%%%%%%%%%%%%%%%%%%%%%%%% 
Figure~\ref{fig:CumulnatsRatio} shows the three cumulant ratios 
\COneTwo, \CThreeTwo, and \CFourTwo\ for the net-$K$, net-$\Lambda$, and 
\netKL\ multiplicity distributions from \sNN=7.7 to 200 GeV. The 
\COneTwo\ for net-kaon and net-$\Lambda$ is always positive as a 
function of centrality. Contrarily, this is negative up to \sNN=39 GeV 
for \netKL. Negative \COne\ is responsible for the negative value of 
this ratio. A similar trend is observed in the case of \CThreeTwo. The 
negative \COne\ and \CThree\ are responsible for making these ratios 
negative for \netKL. The \CFourTwo\ is almost the same for three cases 
at all energies and remains around unity.

On the contrary, the mean value (\COne) of \netKL\ is negative at 
lower energy, whereas it becomes positive at \sNN=200 GeV in Au+Au 
collisions in UrQMD. Here, we reiterate the combined quantity, 
$N_{K+\Lambda}=(N_{K^{+}}+N_{\bar{\Lambda}})-(N_{K^{-}}+N_{\Lambda})$. 
We can understand this trend as the following. More baryons are 
produced at the lower collision energy due to the higher baryon 
deposition. The $\Lambda$ particle, being the lightest strange baryon, 
dominates the \netKL\ and gives rise to this negative value. This 
negative trend decreases as the collision energy increases and 
particle-to-antiparticle yields become similar. This implies more 
negative net-strangeness number (more $s$ quarks) is observed at lower collision 
energies than at the top collision energy. 

Furthermore, to study the correlation contribution between ($N_{K^{+}} 
+ N_{\bar{\Lambda}})$ and ($N_{K^{-}} + N_{\Lambda}$) to \netKL\ 
distribution, we calculate the following expression for the independent 
particle production:
%%%%%%%%%%%%%%%%%%%%%%%%%%%%%%%%%%
\begin{eqnarray}
C_{n}^{K+\Lambda} = C_{n}(N_{K^{+}} + N_{\bar{\Lambda}}) + (-1)^{n} 
C_{n}(N_{K^{-}} + N_{\Lambda}).
\end{eqnarray}
%%%%%%%%%%%%%%%%%%%%%%%%%%%%%%%%%%
It should be noted that the expression mentioned above is built on 
the basis of Eq.~(\ref{Eq:NetKL}). After examining the comparison 
between the assumption of independent particle production and cumulant 
computed from the sample of \netKL\ distributions in the UrQMD model, 
there seems to be a negligible difference. However, it is essential to 
incorporate experimental data to study these correlations in the BES 
energies.

%%%%%%%%%%%%%%%%%%%%%%%%%%%%%%%%%%%%%%%%%%%%%%%%%%%%%%%%%%%%%%%%%%%%%%%%
\begin{figure}[h!]
\centering
\includegraphics[width=0.5\textwidth]{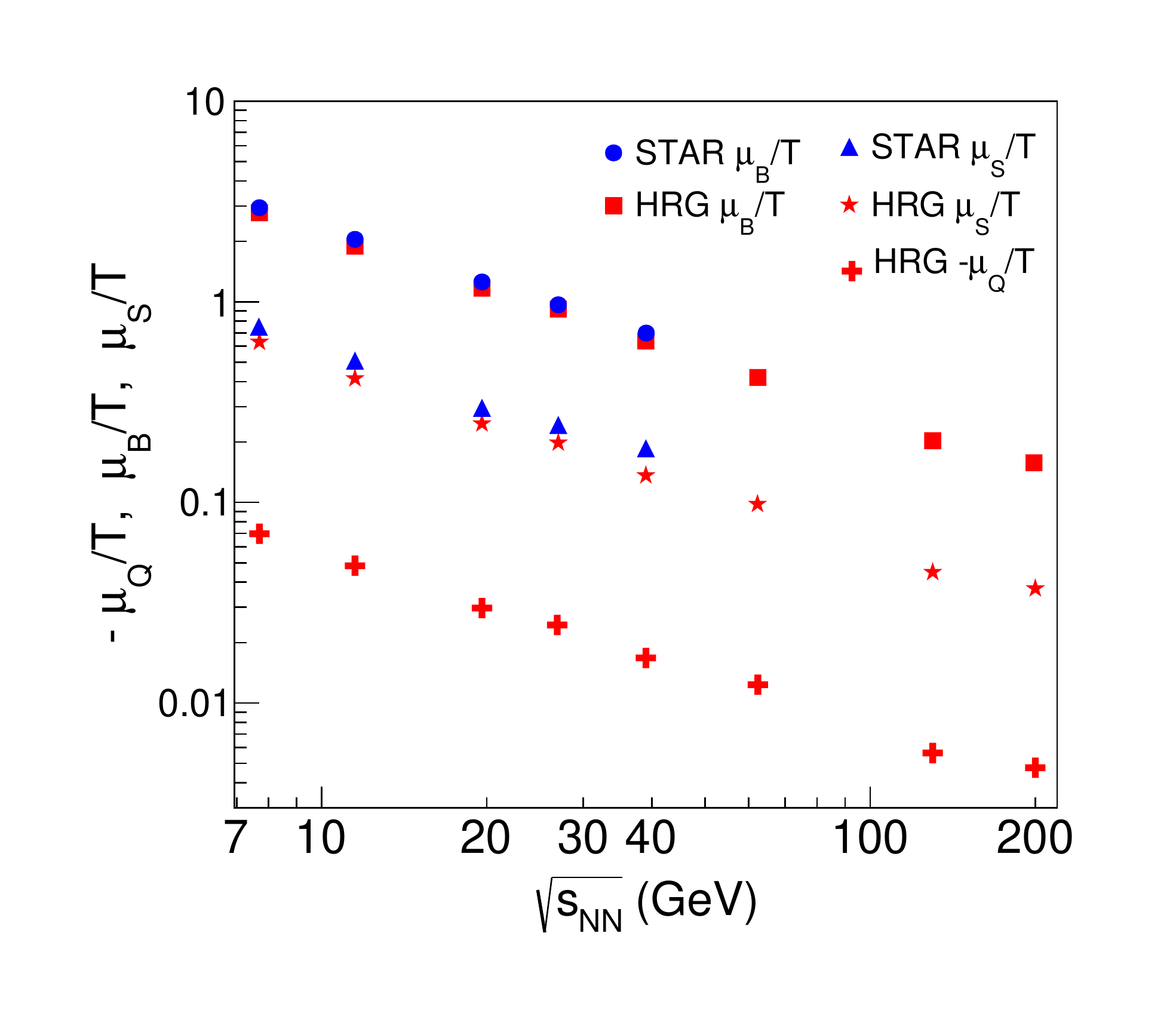}
\caption{The $\mu_B/T$, $\mu_S/T$ and $-\mu_Q/T$ as a 
function of \sNN. The red color markers are the input parameters used 
in HRG calculation to match the STAR net-kaon~\cite{STAR:2017tfy} and 
net-$\Lambda$~\cite{STAR:2020ddh} data as shown in 
Fig.~\ref{fig:chi_Kaon} and ~\ref{fig:chi_Lambda}. The blue markers are the thermal 
model fit from the STAR data~\cite{STAR:2019bjj}.}
 \label{fig:FO}
\end{figure}

%%%%%%%%%%%%%%%%%%%%%%%%%%%%%%%%%%%%%%%%%%%%%%%%%%%%%%%%%%%%%%%%%%%%%%%%
\subsection{Comparison with HRG}
\label{Sect:urqmdHrg}
%%%%%%%%%%%%%%%%%%%%%%%%%%%%%%%%%%%%%%%%%%%%%%%%%%%%%%%%%%%%%%%%%%%%%%%%
A comparison between the UrQMD results and HRG calculation is important 
to understand the underlying hadronic scattering contribution to these 
observables and set a baseline for heavy-ion collision experiments. To 
study the degree of thermalization, we calculated the HRG estimations 
at the chemical freeze-out boundary. Reference \cite{Gupta:2022phu}
has shown that various cumulants (up to third order) of 
different species can describe the susceptibilities calculated in the 
HRG using a grand canonical ensemble above certain colliding energy, 
which restricts the applicability of the HRG model to the fireball 
created in low-energy heavy-ion collisions.

%%%%%%%%%%%%%%%%%%%%%%%%%%%%%%%%%%%%%%%%%%%%%%%%%%%%%%%%%%%%%%%%%%%%%%%%
In Fig.~\ref{fig:FO}, we present the variation of $\mu_B/T$, 
$\mu_S/T$, and $\mu_Q/T$ with collision energy (\sNN), 
which are evaluated by fitting the yield data with the thermal model 
assuming a grand-canonical ensemble. The particle yields, including 
$\pi$, $K$, $p$, $\Lambda$, $\Xi$, and $\Omega$, are used to estimate these 
parameters \cite{Bhattacharyya:2019wag, Bhattacharyya:2019cer, 
Biswas:2020kpu}. While the strangeness and baryon chemical potential 
values are consistent with those reported by the STAR collaboration 
\cite{STAR:2017sal, STAR:2019bjj}, there are no reported values for 
$\mu_Q$. We have included these parameters along with the available 
ones for a comprehensive analysis, as they serve as input parameters 
for HRG calculations. We must emphasize that these parameter values 
result from fitting the yield and are included here for completeness.
%%%%%%%%%%%%%%%%%%%%%%%%%%%%%%%%%%%%%%%%%%%%%%%%%%%%%%%%%%%%%%%%%%%%%%%%
\begin{figure*}[h]
  \centering
\includegraphics[width=1.0\textwidth]{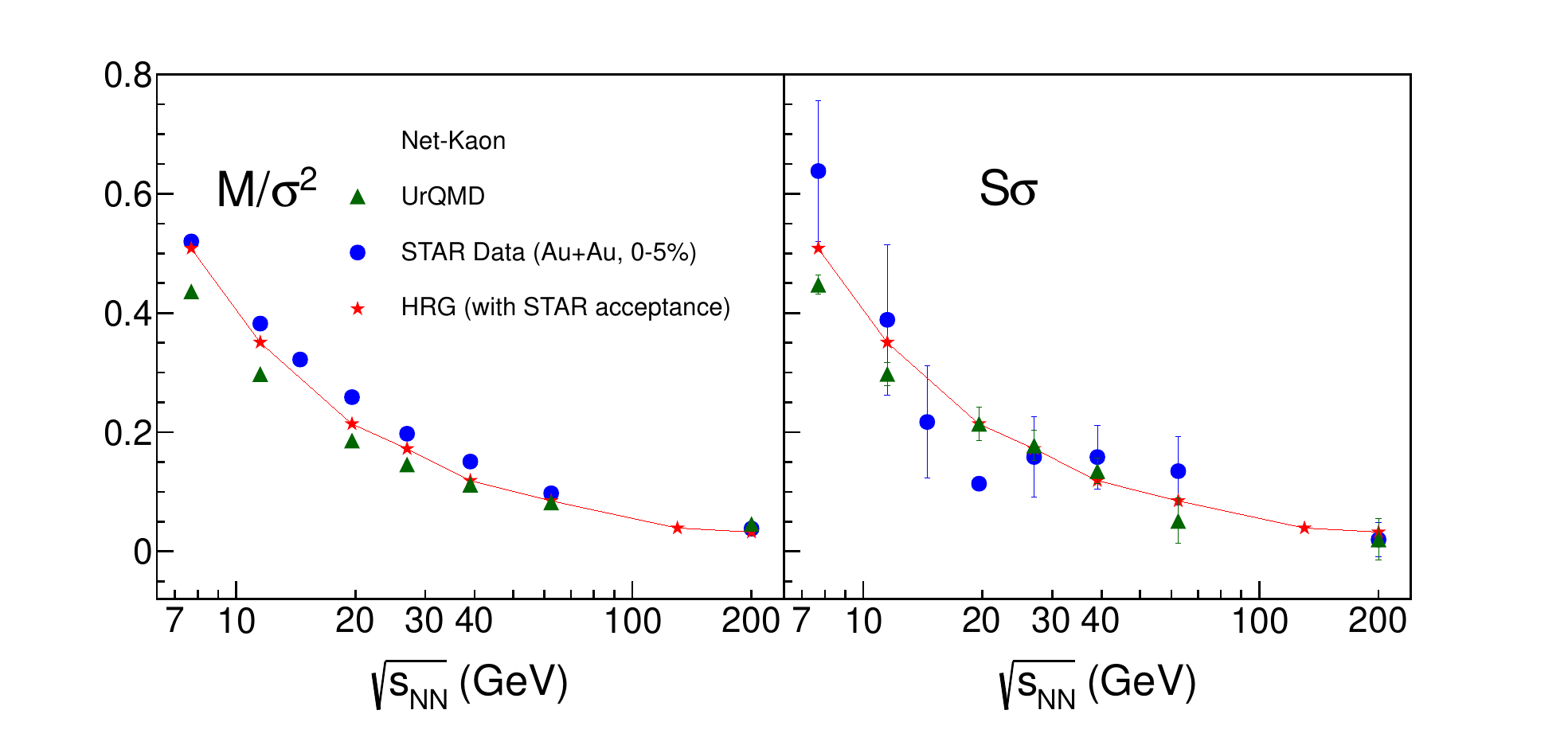} 
\caption{ The $M/\sigma^2$ (\COneTwo) (left panel)  and \Ss $~$ 
(\CThreeTwo) (right panel) of net-$K$ as a function \sNN. The circles, star 
markers, and triangles represent the STAR data, HRG calculations 
(done with the same kinematic cuts as the data), and the UrQMD 
model calculations, respectively. Net-kaon data are from 
Ref. \cite{STAR:2017tfy}.}
\label{fig:chi_Kaon}
\end{figure*}

%%%%%%%%%%%%%%%%%%%%%%%%%%%%%%%%%%%%%%%%%%%%%%%%%%%%%%%%%%%%%%%%%%%%%%%%
Using these parameters, we will initially compare the ratios of 
cumulants for individual species. Previous studies \cite{Garg:2013ata, 
Mishra:2016qyj} have shown that the decay feed-down does not 
significantly affect the ratios of cumulants up to the third 
order for individual species. 

Figure \ref{fig:chi_Kaon} shows the $M/\sigma^2$ and \Ss\ of net-$K$ 
as a function of collision energy (\sNN). The HRG calculations are 
carried out using the same acceptance cuts as employed in the STAR 
measurements \cite{STAR:2020ddh}. The HRG calculation and UrQMD results 
exhibit a good agreement with experimental data for both these ratios. 

%%%%%%%%%%%%%%%%%%%%%%%%%%%%%%%%%%%%%%%%%%%%%%%%%%%%%%%%%%%%%%%%%%%%%%%%
\begin{figure*}[h]
  \centering
\includegraphics[width=1.0\textwidth]{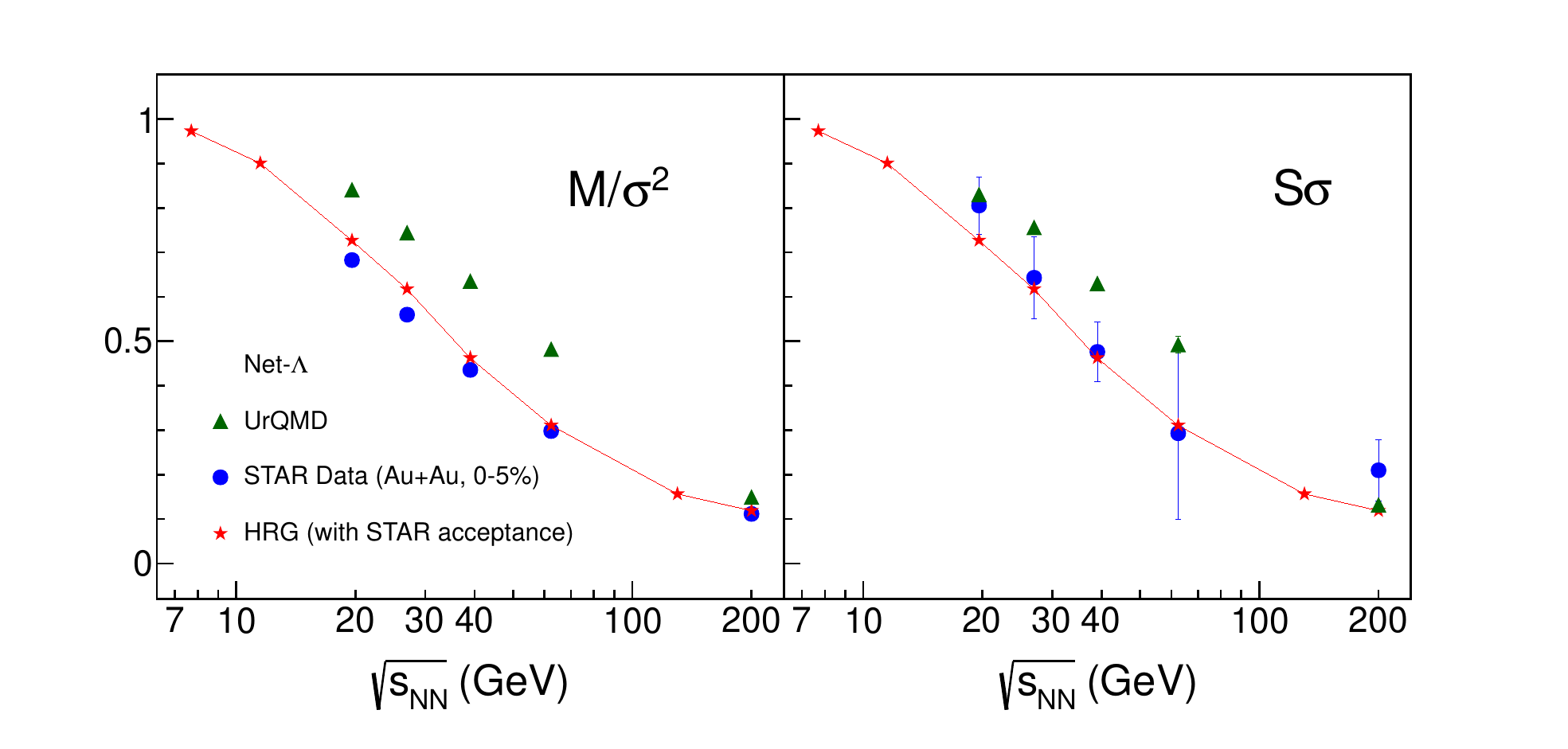} 
\caption{The $M/\sigma^2$ (\COneTwo) (left panel)  and \Ss $~$ 
(\CThreeTwo) (right panel) of net-$\Lambda$ as a function \sNN. The 
circles, star markers, and triangles represent the STAR data, HRG 
calculations (done with the same kinematic cuts as the data), and the UrQMD 
model calculations, respectively. Net-$\Lambda$ data are from 
Ref. \cite{STAR:2020ddh}.}
\label{fig:chi_Lambda}
\end{figure*}

%%%%%%%%%%%%%%%%%%%%%%%%%%%%%%%%%%%%%%%%%%%%%%%%%%%%%%%%%%%%%%%%%%%%%%%%
Figure \ref{fig:chi_Lambda} shows the $M/\sigma^2$ and \Ss\ of net-$\Lambda$ as a 
function of collision energy (\sNN). The HRG calculations well explain 
the data both for the $M/\sigma^2$ and \Ss\  results. The UrQMD 
estimations have a large deviation from the experimental data for 
net-$\Lambda$, which mainly comes from the disagreement in the values of $C_2$ \cite{STAR:2020ddh}. Note here that no decay contributions are 
included in these HRG calculations. 
%The observable \Ss\ predicts thermalization for both species individually. 

%%%%%%%%%%%%%%%%%%%%%%%%%%%%%%%%%%%%%%%%%%%%%%%%%%%%%%%%%%%%%%%%%%%%%%%%
The comparison between the UrQMD model and HRG calculations shows that 
the net-$K$ results are comparable and also match the data. The 
net-$\Lambda$ results have an apparent difference between the UrQMD
and HRG calculations. The UrQMD result for net-$\Lambda$ deviates from 
the data at all energies. The higher mass resonances decay channels are 
included in the UrQMD events. This difference in the net-$\Lambda$ $M/\sigma^2$ and \Ss\ results from the UrQMD could be due to the difference in the $\Lambda$, 
higher resonance (anti)particle production yields in UrQMD, and the 
RHIC energies~\cite{Bass:1998ca}.

Here, it is worth noting that the quantitative similarity between the 
ratios \COneTwo\ ($M/\sigma^2$) and \CThreeTwo\ (\Ss) establishes the 
applicability of the Boltzmann approximation in the context of 
heavy-ion collisions, i.e., $C_3=C_1$. The good agreement among the HRG 
results with the Boltzmann statistics and the experimental data validates 
the Poisson limit of the particle distribution. Any deviation from this 
Poisson limit is due to the additional underlying physics present in the data.
%%%%%%%%%%%%%%%%%%%%%%%%%%%%%%%%%%%%%%%%%%%%%%%%%%%%%%%%%%%%%%%%%%%%%%%%
\begin{figure}[h]
\centering
\includegraphics[width=0.5\textwidth]{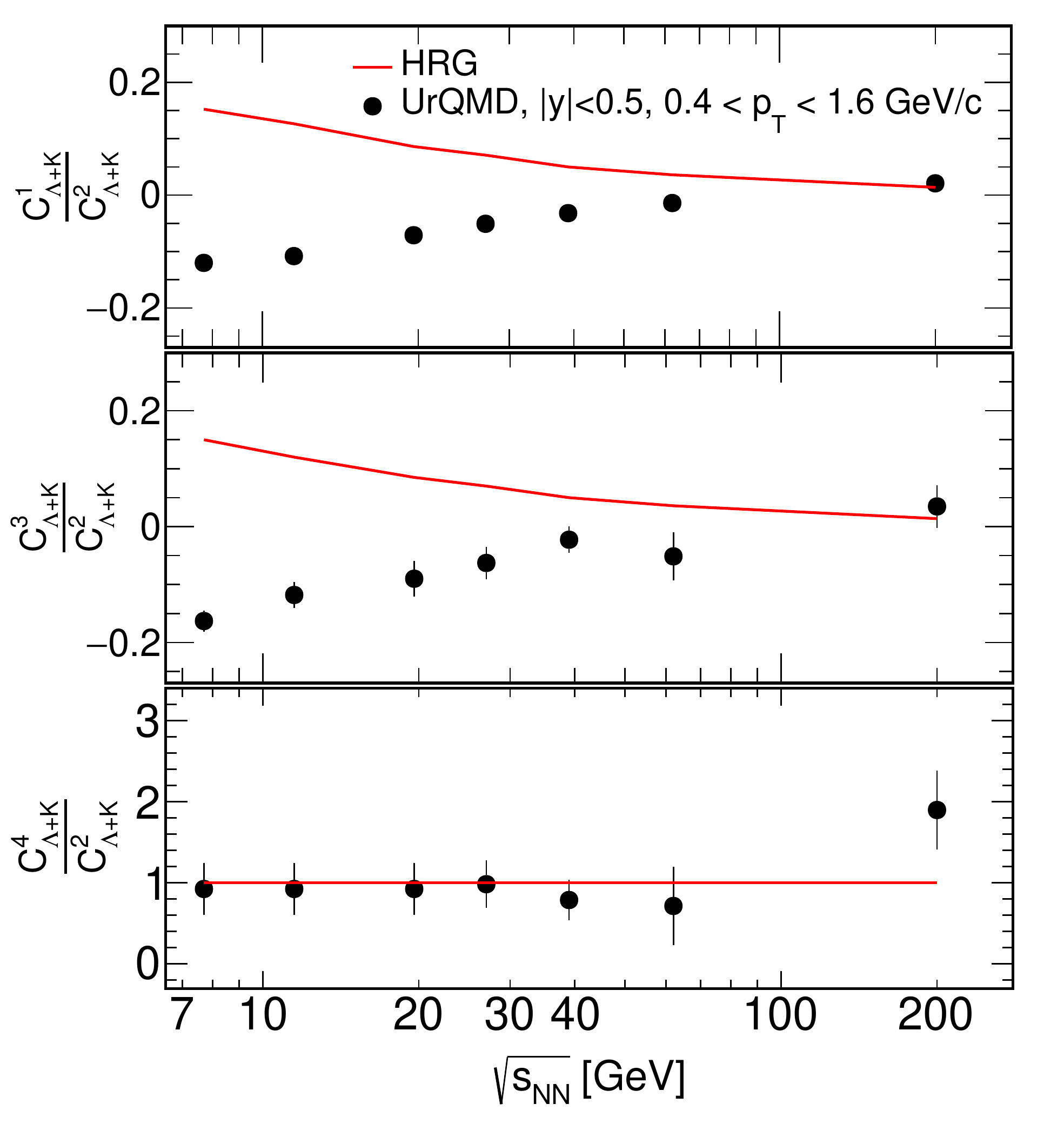}
\caption{The energy dependence of \COneTwo, \CThreeTwo, and \CFourTwo\ 
for 0--5 \% centrality for the UrQMD model (circles) and the HRG 
calculation (red line).}
\label{fig:EngCumulantRatioHrgUrQMD}
\end{figure}
%%%%%%%%%%%%%%%%%%%%%%%%%%%%%%%%%%%%%%%%%%%%%%%%%%%%%%%%%%%%%%%%%%%%%%%%
With a general agreement among UrQMD results, STAR data, and HRG 
calculations, it would be interesting to compare results for the 
\netKL\ multiplicity distribution. As no experimental data are available 
yet for this combination, we have compared the UrQMD results with those 
from the HRG model. In the framework of the ideal HRG model, the 
cumulants of \netKL\ are just the addition of the cumulants of the net-
kaon and net-$\Lambda$. Within the Poisson distribution approximation, 
this simple addition assumes a Skellam distribution, where $C_1=C_3$, 
$C_2=C_4$. The $C_1$ is the difference between the mean in the case of 
cumulants of net quantities. This simplified picture thus provides a 
trivial baseline for the cumulant study. 

In Fig.~\ref{fig:EngCumulantRatioHrgUrQMD}, the 
energy dependence of the \netKL\ \COneTwo, \CThreeTwo, and \CFourTwo\ 
for 0-5\% central Au+Au collisions in the UrQMD model and HRG 
calculations are compared. The \COneTwo\ and \CThreeTwo\ show an 
opposite trend at lower collision energy between the UrQMD and the 
thermal model predictions. The HRG calculations have been performed 
with the chemical freeze-out parametrization from 
Refs. \cite{Bhattacharyya:2019wag, Bhattacharyya:2019cer, 
Biswas:2020kpu}. This disagreement is prominent at lower collision 
energy. This difference arises from higher mass resonance decay in the 
UrQMD model, whereas no such decay feed-down is included in the HRG 
calculations. A detailed discussion on the importance of higher mass 
resonance decay contributions can be found in Sec. 
\ref{Sect:hrgstrngeproduction}. The \CFourTwo\ of HRG calculation is 
unity at all energies and consistent with the UrQMD models. It shows 
that \CFourTwo\ is less sensitive to the resonance contributions.
 
%%%%%%%%%%%%%%%%%%%%%%%%%%%%%%%%%%%%%%%%%%%%%%%%%%%%%%%%%%%%%%%%%%%%%%%%
\subsection{Strangeness production in HRG}
\label{Sect:hrgstrngeproduction}
%%%%%%%%%%%%%%%%%%%%%%%%%%%%%%%%%%%%%%%%%%%%%%%%%%%%%%%%%%%%%%%%%%%%%%%%
In heavy-ion collisions, the strange particles' yields contain the 
contribution from the decay feed-down of higher mass 
resonances~\cite{STAR:2017sal, STAR:2019bjj}. In the thermal model, it 
is important to include all higher mass resonance decay channels to 
capture the bulk description of the chemical freeze-out surface 
properly. Inclusion of these decays for higher order susceptibility 
calculations needs to consider the probabilistic nature of decay 
channels through their branching fractions \cite{Begun:2006jf, 
Nahrgang:2014fza, Xu:2016skm, Mishra:2016qyj}. Including these decay 
products in the yield, calculations are trivial as those are the 
first cumulant (mean). The (anti-)$\Lambda$(1115) has the 
contributions from the higher mass strange baryons, e.g., 
$\Sigma(1385), \Lambda(1405), \Lambda(1520)$, and other 
heavy resonances. Kaons (493) have contributions mostly from the higher 
mass meson resonances like $K^*(892), K(1270), \phi(1020)$, etc. 
However, for kaons, the inclusion of all resonances becomes complex as 
many resonances decay into both $K^+$ and $K^-$. In this work, we have 
not included any decay channels from the higher resonances for the 
susceptibility calculation in the HRG.
%%%%%%%%%%%%%%%%%%%%%%%%%%%%%%%%%%%%%%%%%%%%%%%%%%%%%%%%%%%%%%%%%%%%%%%%
\begin{figure}[h]
\includegraphics[width = 3.5 in,height=3.5 in]{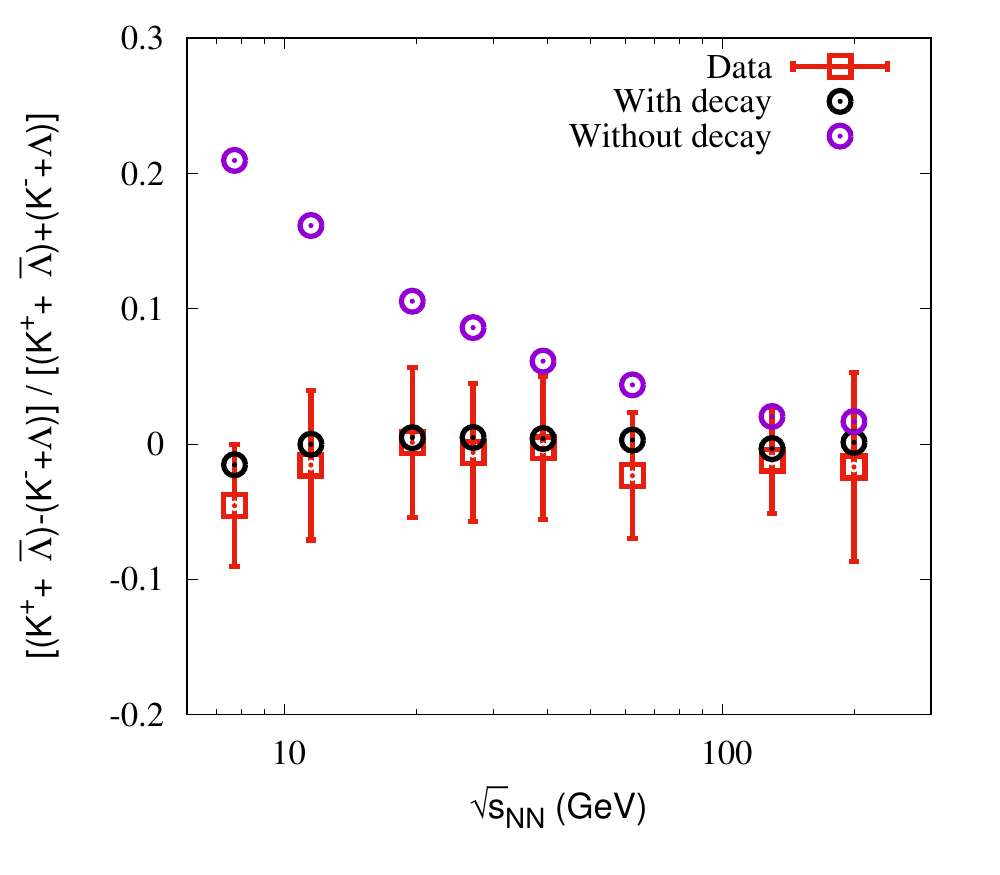}
\caption{The ratio of the net to the total charged kaon and $\Lambda$ 
multiplicity as a function of \sNN. The black  (with decay) and purple 
(without decay) circles represent the HRG calculations.  The data (box) 
are from Ref.\cite{STAR:2019bjj}}
\label{fig:netkl}
\end{figure}
%%%%%%%%%%%%%%%%%%%%%%%%%%%%%%%%%%%%%%%%%%%%%%%%%%%%%%%%%%%%%%%%%%%%%%%%
To quantify the relevance of decay from the higher mass resonances, we 
have plotted the ratio of the mean value of the \netKL\ with that of 
the total in Fig.\ref{fig:netkl}. The model calculations are 
performed following the freeze-out parameters from 
Refs. \cite{Bhattacharyya:2019wag, Biswas:2020dsc, Biswas:2020kpu}. These 
parameters were extracted with the mean value of yields and 
successfully explained the hadronic yield ratios. By including decays 
from higher mass resonances, the model agrees with the data, whereas 
the ratio increases with decreasing beam energy if we exclude the decay 
contribution. Hence, for a reasonable estimate of the parametrization 
and to explain the data, it is necessary to include the higher mass 
resonance decay in the calculation of the net-strangeness observable.

%%%%%%%%%%%%%%%%%%%%%%%%%%%%%%%%%%%%%%%%%%%%%%%%%%%%%%%%%%%%%%%%%%%%%%%%
Although the strangeness neutrality demands the net-strangeness to be 
zero in the total system and distributed among the final particle 
species in the whole phase space. Here, we investigate only the charged 
kaon(493) and $\Lambda$(1115). With increasing $\mu_B$ at lower 
collision energies, strange baryons dominate, and the strangeness gets 
distributed mainly among the hyperons. The $\Lambda$ being the lightest 
one contributes the most significant part. The decay from higher mass 
resonances increases the yield of $\Lambda$ and produces a net negative 
strangeness at lower collision energies. With only the primary 
abundance, the net strangeness remains positive in our observable, as 
the lightest kaons dominate the sum and deliver a positive strangeness. 
Such behaviors can be seen in the UrQMD model calculation as it 
includes all resonance decays in an event, as discussed in 
Sec.~\ref{Sect:urqmdHrg}.

%%%%%%%%%%%%%%%%%%%%%%%%%%%%%%%%%%%%%%%%%%%%%%%%%%%%%%%%%%%%%%%%%%%%%%%%
\section{Summary and outlook}
\label{Sect:summaryoutlook}
%%%%%%%%%%%%%%%%%%%%%%%%%%%%%%%%%%%%%%%%%%%%%%%%%%%%%%%%%%%%%%%%%%%%%%%%
In heavy-ion collision experiments, observables related to net-$K$ and 
net-$\Lambda$ act as a proxy for the strangeness. Although the 
individual results for net-$K$ and net-$\Lambda$ are available from 
STAR BES, results for the combined study for the \netKL\ are yet to be 
performed. In this work, we have studied the cumulants of net-$K$, 
net-$\Lambda$, and \netKL\ multiplicity distributions for the RHIC BES 
energy range using the UrQMD model. These studies serve as a baseline 
for the cumulant measurement of the \netKL\ multiplicity distributions. 

%%%%%%%%%%%%%%%%%%%%%%%%%%%%%%%%%%%%%%%%%%%%%%%%%%%%%%%%%%%%%%%%%%%%%%%%
In UrQMD, the \COneTwo\ and \CThreeTwo\ of \netKL\ are negative at lower 
energies and become positive at higher collision energies within the 
given acceptance window mentioned in this paper. At lower \sNN, the 
finite baryon density favors the dominance of hyperons over strange 
mesons, which produces this negative strangeness. This effect 
diminishes as the collision energy increases and the kaon becomes more 
abundant than the hyperons. On the contrary, the higher-order cumulant 
\CFourTwo\ has no significant variation between net-$K$, net-$\Lambda$, and 
\netKL\ multiplicity distributions. 

%%%%%%%%%%%%%%%%%%%%%%%%%%%%%%%%%%%%%%%%%%%%%%%%%%%%%%%%%%%%%%%%%%%%%%%%
As a benchmark, we have compared our UrQMD calculations of various 
cumulants with available STAR data of net-$K$ and net-$\Lambda$ for the 
most central collision, where a good agreement is apparent. 
Furthermore, we have compared the UrQMD results with the HRG 
calculation to study the thermalization contribution to these 
observables. The HRG calculations have been performed at the standard 
chemical freeze-out parametrization. Although there is good agreement 
among data, UrQMD calculations and model(HRG) predictions for 
the individual net-$K$ and net-$\Lambda$, differences among UrQMD 
results and HRG for cumulant ratios of \netKL\ are apparent. 

%%%%%%%%%%%%%%%%%%%%%%%%%%%%%%%%%%%%%%%%%%%%%%%%%%%%%%%%%%%%%%%%%%%%%%%%
It seems that the difference between the decay feed-down of higher mass 
resonances is responsible for the difference between UrQMD and HRG. The 
decay consideration is necessary to explain the available experimental 
data, which is responsible for a negative \COne\ and \CThree\ at lower 
collision energies.

%%%%%%%%%%%%%%%%%%%%%%%%%%%%%%%%%%%%%%%%%%%%%%%%%%%%%%%%%%%%%%%%%%%%%%%%
This \netKL\ cumulant measurements along with that of net-$K$ and 
net-$\Lambda$ multiplicity distributions can provide necessary 
information on the strangeness in heavy-ion collisions at RHIC 
energies, and will act as a benchmark for future experiments. It is important to compare these calculations from the UrQMD 
and HRG with the STAR ongoing measurements. The STAR's BES-II data with 
lower collision energies in the fixed target experiment could provide 
important information about the strangeness production and their event-
by-event fluctuations in heavy-ion collisions. The proper treatment of 
the decay feed-down into kaon and $\Lambda$ would facilitate the 
extraction of the chemical freeze-out parameters from the strangeness 
sector.

%%%%%%%%%%%%%%%%%%%%%%%%%%%%%%%%%%%%%%%%%%%%%%%%%%%%%%%%%%%%%%%%%%%%%%%%
\section*{Acknowledgement}
We would like to thank Bedangadas Mohanty, Tapan Nayak, Sayantan Sharma, 
Qinghua Xu, Zhangbu Xu, and Li Yi for their fruitful discussion and 
comments on this work. N.R.S. would like to acknowledge IMSc, Chennai, for 
hospitality during his visit where this work was started.  
N.R.S. is supported by the Fundamental Research Funds of Shandong 
University and the National Natural Science Foundation of China, Grant 
No. 12050410235.

%\newpage

\end{document}